\documentclass[12pt]{article}
\usepackage{fullpage,citesort,epsfig,psfrag,graphics,amsbsy,amssymb}
\usepackage{caption}
\newcommand{\beq}{\begin{equation}}
\newcommand{\eeq}{\end{equation}}
\newcommand{\bea}{\begin{eqnarray}}
\newcommand{\eea}{\end{eqnarray}}

\newcommand{\Tr}{{\rm Tr}}
\newcommand{\be}{\begin{equation}}
\newcommand{\ee}{\end{equation}}
\newcommand{\bq}{\begin{eqnarray}}
\newcommand{\eq}{\end{eqnarray}}
\newcommand{\ket}[1]{|#1\rangle}
\newcommand{\bra}[1]{\langle#1|}

\def\math{\mathsurround=0pt }
\def\leftrightarrowfill{$\math \mathord\leftarrow \mkern-6mu 
 \cleaders\hbox{$\mkern-2mu \mathord- \mkern-2mu$}\hfill
 \mkern-6mu \mathord\rightarrow$}
\def\overleftrightarrow#1{\vbox{\ialign{##\crcr
     \leftrightarrowfill\crcr\noalign{\kern-1pt\nointerlineskip}
     $\hfil\displaystyle{#1}\hfil$\crcr}}}

\newcommand{\VEV}[1]{\langle#1\rangle}

\let\l=\lambda

 \def\bd{\begin{document}} \def\ed{\end{document}}
\def\ds{\documentstyle} \let\fr=\frac \let\bl=\bigl \let\br=\bigr
\let\Br=\Bigr \let\Bl=\Bigl
\let\bm=\bibitem
\let\na=\nabla
\let\pa=\partial \let\ov=\overline
\def\ft#1#2{{\textstyle{{\scriptstyle #1}\over {\scriptstyle #2}}}}
\def\fft#1#2{{#1 \over #2}}
\def\vp{\varphi}
\def\sst#1{{\scriptscriptstyle #1}}
\def\oneone{\rlap 1\mkern4mu{\rm l}}
\def\td{\tilde}
\def\wtd{\widetilde}
\def\dalemb#1#2{{\vbox{\hrule height .#2pt
        \hbox{\vrule width.#2pt height#1pt \kern#1pt
                \vrule width.#2pt}
        \hrule height.#2pt}}}
\def\square{\mathord{\dalemb{6.8}{7}\hbox{\hskip1pt}}}
\def\wtd{\widetilde}
\def\R{\rlap{\rm I}\mkern3mu{\rm R}}
\def\im{{\rm i}}
\def\tilg{\tilde{g}}
\def\tilF{\tilde{F}}
\def\tilA{\tilde{A}}
\def\varf{\varphi}
\def\tilf{\tilde{\phi}}
\def\tilh{\tilde{h}}
\def\rme{{\rm e}}
\def\ep{\epsilon}
\def\0{{(0)}}
\def\9{{(9)}}
\def\8{{(8)}}
\def\7{{(7)}}
\def\6{{(6)}}
\def\5{{(5)}}
\def\4{{(4)}}
\def\3{{(3)}}
\def\2{{(2)}}
\def\1{{(1)}}
\newcommand{\trace}{{\rm Tr}}
\newcommand{\ub}{\overline{U}}
\newcommand{\vb}{\overline{V}}
\newcommand{\uh}{\widehat{U}}
\newcommand{\vh}{\widehat{V}}
\newcommand{\ubh}{\overline{\widehat{U}}}
\newcommand{\vbh}{\overline{\widehat{V}}}
\newcommand{\lb}{\bar{\l}}
\newcommand{\Fb}{\overline{F}}
\newcommand{\Fh}{\widehat{F}}
\newcommand{\Fbh}{\overline{\widehat{F}}}
\newcommand{\Ab}{\overline{A}}
\newcommand{\Ah}{\widehat{A}}
\newcommand{\Abh}{\overline{\widehat{A}}}
\newcommand{\Gb}{\overline{G}}
\newcommand{\Gh}{\widehat{G}}
\newcommand{\Gbh}{\overline{\widehat{G}}}
\newcommand{\Pb}{\overline{P}}
\newcommand{\Ph}{\widehat{P}}
\newcommand{\Pbh}{\overline{\widehat{P}}}
\newcommand{\Qb}{\overline{Q}}
\newcommand{\Qh}{\widehat{Q}}
\newcommand{\Qbh}{\overline{\widehat{Q}}}
\newcommand{\Bb}{\overline{B}}
\newcommand{\Bh}{\widehat{B}}
\newcommand{\Bbh}{\overline{\widehat{B}}}
\newcommand{\fhns}{\hat{F}^{\rm (NS)}}
\newcommand{\fhrr}{\hat{F}^{\rm (RR)}}
\newcommand{\ahns}{\hat{A}^{\rm (NS)}}
\newcommand{\ahrr}{\hat{A}^{\rm (RR)}}
\newcommand{\hhrr}{\hat{H}^{\rm (RR)}}
\newcommand{\hchi}{\hat{\chi}}
\newcommand{\hphi}{\hat{\phi}}
\newcommand{\htau}{\hat{\tau}}
\newcommand{\cG}{{\cal G}}
\newcommand{\cGb}{\overline{{\cal G}}}
\newcommand{\cH}{{\cal H}}
\newcommand{\cP}{{\cal P}}
\newcommand{\cPb}{\overline{{\cal P}}}
\newcommand{\cQ}{{\cal Q}}
\newcommand{\cQb}{\overline{{\cal Q}}}
\newcommand{\cM}{{\cal M}}
\newcommand{\cN}{{\cal N}}
\newcommand{\cO}{{\cal O}}
\newcommand{\cD}{{\cal D}}
\newcommand{\cL}{{\cal L}}

\newcommand{\vpp}{\mbox{$\langle{\scriptstyle++}\rangle$}}
\newcommand{\vmp}{\mbox{$\langle{\scriptstyle-+}\rangle$}}
\newcommand{\vppp}{\mbox{$\langle{\scriptstyle+++}\rangle$}}
\newcommand{\vmpp}{\mbox{$\langle{\scriptstyle-++}\rangle$}}
\newcommand{\vpmp}{\mbox{$\langle{\scriptstyle+-+}\rangle$}}

\begin{document}
\setlength{\captionmargin}{36pt}
\begin{titlepage}
\begin{flushright}
UFIFT-HEP-08-15\\
\end{flushright}

\vskip 3cm
\begin{center}
\begin{large}
{\bf Nonabelian D-branes, Open Strings, and Gauge Theory}
\footnote{Supported in part by the Department
of Energy under Grant No. DE-FG02-97ER-41029.} 
\end{large}

\vskip 2cm
{\large 
Charles B. Thorn\footnote{E-mail  address: {\tt thorn@phys.ufl.edu}}
}
\vskip0.20cm
{\it Institute for Fundamental Theory\\
Department of Physics, University of Florida,
Gainesville FL 32611}


\vskip 1.0cm
\end{center}

\begin{abstract}
\noindent There is a subtle difference between the open string 
dynamics determined by the original dual resonance models
and that determined by D-brane constructions within
critical closed string theory. For instance, in contrast
to the former, the latter have
massless scalars in addition to the massless gluon shared by both. 
We introduce and explain the
concept of nonabelian D-branes which illuminates this distinction.
We employ this concept to offer new possibilities for string duals
for large N QCD (pure Yang-Mills gauge theory). 
\end{abstract}
\vfill
\end{titlepage}
\section{Introduction}
In the decade since Maldacena proposed the AdS/CFT correspondence,
establishing the equivalence between ${\cal N}=4$ supersymmetric Yang-Mills
in 4 spacetime dimensions and type IIB superstring theory on an
AdS$_5\times$S$^5$ background \cite{maldacenasole}, 
it has been a major challenge to
find a string dual for QCD. Most attempts in this direction
take the conformal  ${\cal N}=4$ case as a starting point and seek to
break the symmetries down to those of QCD. In 't Hooft's
 $N\to\infty$ limit \cite{thooftlargen},
which only involves the gluonic sector of QCD, 
this requires giving large masses to all gauginos and
to the 6 unwanted scalars on the field side of the duality, 
with a corresponding modification of the background on the string side
\cite{wittenadsblackhole}.
To describe QCD at finite $N$ one must also 
add new degrees of freedom to describe
quarks \cite{karchkatz}.

A more economical approach, which has neither gauginos nor the 6 unwanted
scalars from the start, is to base the string dual for QCD on
subcritical string theory \cite{polyakov,curtrightthorn,gervaisneveu}
in 4 space-time dimensions.
Inspired by \cite{thornsubdual,preitschopf,polyakovwall}, 
we proposed the even G-parity
sector \cite{mandelstam,gliozziso}
of the Neveu-Schwarz open string model 
in 4 spacetime dimensions \cite{neveuschwarz,neveust}  
as the natural starting point \cite{thornsubqcd}. 
For brevity we call this the NS+ model. 
Its low energy limit ($\alpha^\prime\to0$) is, in perturbation theory, 
precisely pure Yang-Mills gauge theory with no extra scalar 
fields and no tachyon \cite{neveuscherk}. 
It is probably the simplest open
string model with this property. The
idea for reaching large N QCD 
is to take the limit $\alpha^\prime\to0$ only after summing
over all planar open string multi-loop diagrams. The hope is that
the graph summation is more tractable in string theory than in
field theory, because it can be interpreted as a tree-level
shift in the closed string background.

The biggest technical challenge with this subcritical approach is that the
perturbative closed string, interacting with the subcritical open
string, is imperfectly understood, because it is so  qualitatively
different from all known (critical) closed string systems. 
One striking difference, seen in the open string nonplanar one loop diagram, is
that the would be ``graviton'', the lightest spin two state
of the closed string, is massive. All known closed string
tree amplitudes predict this state to be massless. Thus a
new perturbative closed string theory, determined in principle
by the multi-loop open string amplitudes, must be discovered and
developed. This is a very interesting (and daunting) problem, which we do not
address here.

Instead, our goal is to clarify and then exploit a subtle difference 
between what we call the true open string and the open string
obtained as the T-dual of normal D-brane constructions \cite{dailp} within
critical closed string theory.  
The T-dual description of an open string with free ends, is an
open string with its ends fixed to D-branes. In this way one can
understand open strings as fluctuations of a closed string background.
To describe a 4 dimensional field theory one introduces a stack of
D3-branes on which open strings end. The spectrum of these open
strings includes 4D massless vectors associated with
open string vibrations parallel to the branes,
as well as massless scalars corresponding to moduli for
fluctuations of the
branes into the extra dimensions transverse to the branes. In the
26 dimensional
bosonic string model there would be 22 such massless scalars, whereas in
the 10 dimensional NS or NS+ model there would be 6. 
The string dual to QCD cannot
have any extra massless scalars. The fact that these extra massless
scalars are not present in true open string theory in 4 dimensions
motivated the proposal of \cite{thornsubqcd}. 

The easiest way to understand this difference is to compare
a 24-brane in the 26 dimensional bosonic string theory
or an 8-brane in the 10 dimensional NS+ model with the open
string in 25, 9 dimensions respectively. The massless level
of 25 (9) dimensional bosonic (NS+) open string theory has 
only 23 (7) degrees of freedom (as a 25 (9) dimensional gluon should), 
whereas in the brane incarnation there are 24 (8) massless degrees
of freedom, the 24th (8th) being a modulus for string fluctuations
transverse to the brane. In particular, the usual
brane realization of $D=25$ open string has one extra
scalar adjoint field in its low energy limit, which
is absent in the low energy limit of the true open string theory.
This is the reason why the low energy limit of open string models
is pure Yang-Mills gauge theory, while in the D-brane interpretation
the low energy limit has additional matter degrees of freedom.

In this paper we introduce the concept of nonabelian D-branes
to describe true open strings as fluctuations of
a closed string background. The usual (abelian) D-brane
conditions only restrict the zero mode $p_0$, of the open string
excitations transverse to the brane, to $p_0=0$. We propose a stronger
nonabelian constraint that also knocks out the lowest non-zero
modes ($a_{\pm1}, b_{\pm1/2}$). We construct operators $J_\pm$
which commute with the (super) Virasoro algebra, in terms of which
the nonabelian D-brane constraints are
\bea
J_\pm\ket{\rm Phys}=0,\qquad  p_0\ket{\rm Phys}=0
\eea
When the dimension of space-time is one less than critical, that
is when $D=25 (9)$,
$J_3\propto p_0$, $J_\pm\equiv J_1\pm i J_2$, enjoy the
Lie algebra of SU(2) (O(3)).

We develop these ideas in stages. In Section 2 we discuss
the nonabelian D-brane construction which describes 
the $D=25$ open bosonic string. Section 3 extends the 
construction to the $D=9$ open NS+ string, and Section 4 discusses
general subcritical $D$. Finally in Section 5 we discuss
various new options for a string dual of large N QCD.

\section{Open bosonic string, $D=25$}
When the physical state conditions are solved in $D<26$ dimensional
(subcritical) open string theory \cite{browert}, 
the null space is (relatively) smaller
than in $D=26$ \cite{brower,goddardt}.
In the latter case the null space is so big that fully two
components of each oscillator decouple. This counting is reflected in the
partition function
\bea
\Tr_{\rm Phys} w^R = \prod_{n=1}^\infty (1-w^n)^{-24},\qquad {\rm for}~D=26,
\label{openpart26}
\eea
where $R=\sum_{n=1}^\infty a_{-n}\cdot a_n$ is the (mass)$^2$ level operator
of bosonic open string theory.
In contrast, for $D<26$ one has
\bea
\Tr_{\rm Phys} w^R =(1-w)\prod_{n=1}^\infty (1-w^n)^{-(D-1)}
,\qquad {\rm for}~D<26.
\label{openpartd}
\eea
For $D=25$ this last reads
\bea
\Tr_{\rm Phys} w^R =(1-w)\prod_{n=1}^\infty (1-w^n)^{-24}
,\qquad {\rm for}~D=25.
\label{openpart25}
\eea
In the D-brane interpretation of open strings, the ends
of an open string are fixed to a D-brane, but its interior
is allowed to vibrate in the full 26 dimensional space-time.
For example, taking a 24-brane, its associated 25 dimensional open string
would have a partition function identical to (\ref{openpart26})
{\it not} (\ref{openpart25}). The extra factor of $(1-w)$ in the
latter means, among other things, 
that the massless first excited open string state has 23
not 24 degrees of freedom. There is an extra massless scalar
in the D-brane interpretation, compared to the open string model. 

We ask therefore whether a more sophisticated brane construction
can describe the true open string theory, without this extra
massless scalar.
For the bosonic string such a construction, which we briefly review in the
following, has in essence been given
before \cite{thornsubdual,preitschopf} albeit in a different context. 
Consider a 24-brane at $x^{25}=0$ in critical bosonic
string theory. 
Then open strings are required to satisfy $x^{25}({\rm ends})=0$.
It is convenient to do a T-duality transform from $x^{25}$
to $y^{25}(\sigma,\tau)$ satisfying Neumann conditions. Then the
Dirichlet conditions on $x^{25}$ translate to the condition that
$y^{25}$ carries zero momentum $p_0^{25}=0$. It is as though
$y^{25}$ had been compactified and only the zero momentum mode
kept. Had we kept all momentum modes after compactification of
$y^{25}$ to a circle, the
possible values of open string momenta would be $p_0=2\pi n/r$.
For the special case $r=4\pi\sqrt{\alpha^\prime}$ it is known
\cite{kacmoody} that the full spectrum of this $c=1$ conformal
field theory supports
 an SU(2) symmetry. At the level of spectrum this can be seen
by first identifying the SU(2) generator $J_3=p^{25}_0\sqrt{\alpha^\prime}$
and then considering the SU(2) character of a single free
boson including compactified momentum modes, $L_0=R+\alpha^\prime p_0^2$,
\bea
\Tr w^{L_0}e^{-i\theta J_3}=
\sum_{k=-\infty}^\infty w^{k^2/4}e^{ik\theta/2}
\prod_{n=1}^\infty{1\over1-w^n}\ .
\eea
The SU(2) character for irreducible representation $j$ is
$\chi_j(\theta)=\sum_{m=-j}^je^{im\theta}$ so we can
write $e^{ik\theta/2}+e^{-ik\theta/2}=\chi_{k/2}-\chi_{(k-2)/2}$
from which we see
\bea
\Tr w^{L_0}e^{-i\theta J_3}=
\sum_{k=0}^\infty w^{k^2/4}\chi_{k/2}(\theta)(1-w^{k+1})
\prod_{n=1}^\infty{1\over1-w^n}=\sum_{k=0}^\infty \chi_{k/2}\
w^{k^2/4}\prod_{n\neq k+1}
{1\over1-w^n}\ .
\eea
This shows that the spectrum supports an SU(2) symmetry, because
each irreducible representation occurs a positive number of times.
 Furthermore, it shows that the SU(2) invariant subspace, $k=0$, has the
partition function
\bea
\prod_{n=2}^\infty{1\over1-w^n}=(1-w)\prod_{n=1}^\infty{1\over1-w^n}\ ,
\eea
precisely the counting of longitudinal states needed to give the
$D=25$ open string.
In this language the usual brane construction gives the
$J_3=0$ subspace, the subspace invariant under rotations
about the 3-axis. By strengthening the restriction to the
full nonabelian SU(2) invariance we get the true open string
spectrum. We call this a nonabelian D-brane.

There is also an explicit construction of the SU(2) generators,
$J_3,J_{\pm}=J_1\pm iJ_2$, based on vertex operators 
\cite{kacmoody}:
\bea
J_{\pm}&=&\oint {dz\over2\pi i z}:e^{\pm iy^{25}(z)/\sqrt{\alpha^\prime}}:,
\qquad J_3=\sqrt{\alpha^\prime}p^{25}_0
\eea
Because the vertex operator $:~e^{\pm iy^{25}(z)/\sqrt{\alpha^\prime}}~:$
has conformal weight $1$, we have $[L_n,J_\pm]=0$
(that $[L_n,J_3]=0$ is trivial) provided the
state space is restricted to states 
with momenta $p^{25}_0=\mathbb{Z}/2\sqrt{\alpha^\prime}$, 
i.e. the state space of
the compactified $y^{25}$. Thus the SU(2) invariance constraint
can be consistently applied to the physical subspace, and we can
then characterize the nonabelian D-brane physical open string states
by the succinct conditions
\bea
(L_0-1)\ket{{\rm Phys}}=L_n\ket{{\rm Phys}}=J^a\ket{{\rm Phys}}=0,
\qquad n>0,\quad a=1,2,3.
\eea
Here the $L_n$ contain all 26 components of the $a_n, p_0$ operators.
Note that the ``normal'' D-brane condition $J_3=\sqrt{\alpha^\prime}
p_0=0$ ensures that the $J_\pm$ are properly defined on the
physical subspace. The non-null solutions of these conditions are in 1-1
correspondence with the non-null physical states of the
open bosonic string in 25 space-time dimensions.

The SU(2) invariant subspace of the Fock space of $y^{25}$ can 
also be identified with the conformal block of the $c=1$ 
Virasoro generators $L^{25}_n\equiv \ell_n$, built on the
$p_0^{25}=0$ primary state $\ket{0,0}$, for which it
is immediate that $J_a\ket{0,0}=0$. Because it has
zero momentum it also satisfies $\ell_{-1}\ket{0,0}=0$, so this
conformal block is spanned by the states
\bea
\ell_{-n}^{\lambda_n}\cdots 
\ell_{-3}^{\lambda_3}\ell_{-2}^{\lambda_2}\ket{0,0},
\eea
which are all SU(2) invariant because $[J^a,\ell_{-n}^{25}]=0$.
The partition function for this subspace is evidently
$\prod_{n=2}^\infty(1-w^n)^{-1}$, precisely as needed for the longitudinal
open string physical states.

\section{Open Neveu-Schwarz string, D=9}
The NS open string in $D<10$ 
space-time dimensions \cite{neveuschwarz,neveust} 
has physical states, with
counting analogous to the bosonic string, described by 
\cite{goddardw,goddardt} the partition functions\footnote{
Here we use the Picture 2 Fock space \cite{neveust} in
which the massless gluon state is
$b_{-1/2}\ket{0,p}$.}
\bea
\Tr_{\rm Phys} w^R(\pm)^{2R} =(1\mp w^{1/2})\prod_{r=1/2}^\infty
(1\pm w^r)^{(D-1)}\prod_{n=1}^\infty (1-w^n)^{-(D-1)}
,\qquad {\rm NS~ for}~D<10.
\label{nsopenpartd}
\eea
The $(-)$ case is needed in constructing the 
projector $(1-(-)^{2R})/2$ onto even G-parity. 
Here $R=\sum_{n=1}^\infty a_{-n}\cdot a_n+\sum_{r=1/2}^\infty
rb_{-r}\cdot b_r$.
As in the bosonic theory the extra factor $(1-w^{1/2})$ accounts for
the transversality of the massless gluon state.
For the critical dimension
\bea
\Tr_{\rm Phys} w^R(\pm)^{2R} &=&\prod_{r=1/2}^\infty
(1\pm w^r)^{(D-2)}\prod_{n=1}^\infty (1-w^n)^{-(D-2)}\nonumber\\
&\to&\prod_{r=1/2}^\infty
(1\pm w^r)^{8}\prod_{n=1}^\infty (1-w^n)^{-8}
,\qquad {\rm NS~ for}~D=10.
\label{nsopenpart10}
\eea
Again for $D=9$ (one less than critical) the counting is not quite
the same as critical:
\bea
\Tr_{\rm Phys} w^R(\pm)^{2R} &=&(1\mp w^{1/2})\prod_{r=1/2}^\infty
(1\pm w^r)^{8}\prod_{n=1}^\infty (1-w^n)^{-8}\nonumber\\
&=&\left({1\pm w^{1/2}\over 1-w}\right)^7\prod_{r=3/2}^\infty
(1\pm w^r)^{8}\prod_{n=2}^\infty (1-w^n)^{-8}
,\qquad {\rm NS~ for}~D=9,
\label{nsopenpart9}
\eea
differing in the extra factor $1\mp w^{1/2}=(1-w)/(1\pm w^{1/2})$ implying that
the gluon state has only 7 degrees of freedom. For a normal
8-brane in 10 space-time dimensions this state would have
8 degrees of freedom. 

So we should  try to interpret the NS
open string in 9 space-time dimensions as a nonabelian
8-brane. Accordingly, we impose $x^9({\rm ends})=0$, 
do a T-duality transformation
to $y^9$ with momentum constraint $p_0^9=0$, and seek
a nonabelian generalization of this condition. The nonabelian
generators must commute with the super Virasoro algebra
$G_r, L_n$, so we consider the integrated vertex operators
$H(z):e^{\pm iy^9(z)/\sqrt{2\alpha^\prime}}:$, which have conformal
weight~1,
\bea
J_{\pm}=\sqrt{2}\oint {dz\over2\pi i z}H(z)
:e^{\pm iy^9(z)/\sqrt{2\alpha^\prime}}:
\ .
\eea
One might think these are fermionic operators, but they are actually
bosonic because $:~e^{iy^9(z)/\sqrt{2\alpha^\prime}}~:$
is just a bosonized fermionic field when acting on the
states with $p_0^9\in\mathbb{Z}/\sqrt{2\alpha^\prime}$. The $J_\pm$
commute with both $G_r$ and $L_n$ on this subspace of states.
If we identify $J_3=p_0\sqrt{2\alpha^\prime}$, so that
$[J_3,J_\pm]=J_\pm$, we then find, with the above normalization, that
$[J_+,J_-]=2J_3$. The SU(2) character on this space of states
is
\bea
\Tr w^{L_0}(\pm)^{2R}e^{-i\theta J_3}&=&
\sum_{k=-\infty}^\infty w^{k^2/2}e^{ik\theta}
\prod_{r=1/2}^\infty(1\pm w^r)\prod_{n=1}^\infty{1\over1-w^n}\\
&=&\sum_{k=0}^\infty w^{k^2/2}\chi_k(\theta)(1\mp w^{k+1/2})
\prod_{r=1/2}^\infty(1\pm w^r)\prod_{n=1}^\infty{1\over1-w^n}\\
&=&\sum_{k=0}^\infty w^{k^2/2}\chi_k(\theta)
\prod_{r\neq k+1/2}^\infty(1\pm w^r)\prod_{n\neq2k+1}^\infty{1\over1-w^n}
\eea
In the NS model we see that spinor representations of SU(2)
are absent, so it is more accurate to say that the symmetry is
O(3). This formula shows that the partition function for
the O(3) invariant subspace is
\bea
(1\mp w^{1/2})\prod_{r=1/2}^\infty(1\pm w^r)\prod_{n=1}^\infty{1\over1-w^n}
=\prod_{r=3/2}^\infty(1\pm w^r)\prod_{n=2}^\infty{1\over1-w^n},
\label{nspart}
\eea
which enumerates precisely the number of longitudinal states
needed to describe the NS open string in 9 space-time dimensions.
Again these states are in 1-1 correspondence with the 
superconformal block  built on the O(3) invariant
primary $\ket{0,0}$:
\bea
\ell_{-n}^{\lambda_n}\cdots\ell_{-2}^{\lambda_2}g_{-r}^{\gamma_r}
\cdots g_{-3/2}^{\gamma_{3/2}}\ket{0,0}
\eea
where $\ell_n,g_r$ are the $c=1$ super-Virasoro generators
built out of $a^9_{-n}, b^9_{-r}$. Factors of $\ell_{-1}, g_{-1/2}$
are absent because they kill $\ket{0,0}$ since $a^9_0\ket{0,0}=0$.

Finally we mention that we could also describe this $c=1$ system
in a manifestly O(3) invariant way by
replacing the bosonic $y^9$ with a pair of fermion fields $H^1, H^2$
and by identifying the original $H$ as $H^3$. Then the three $H^a$
transform as a vector under $O(3)$ with generators 
\bea
J_a=\epsilon_{abc}
\oint {dz\over2\pi i z}H^b(z)H^c(z).   
\eea
Then the nonabelian D-brane condition would be $J_a\ket{\rm Phys}=0$.

\section{General $D$}
For the nonabelian D-brane construction to work for the open
bosonic string in 25 space-time dimensions and the open
NS string in 9 space-time dimensions, it was essential that the
physical state conditions effectively removed {\it two} components
from each mode. This will be the case if the (super)Virasoro
central charge $c$ is critical ($c=26, 10$ respectively). In each case,
adding the extra dimension promotes the $c$ to critical. 
Imposing SU(2) (O(3))
invariance knocks out the first mode of the extra dimension
and finally the critical (super) Virasoro conditions remove
two components leading to (\ref{openpart25}) or (\ref{nsopenpart9}) 
respectively.

This same construction doesn't immediately work for other subcritical
dimensions, because then adding the extra dimension only promotes
$c=D$ to $c=D+1$ and, for $D+1<$ critical, the physical state conditions
only remove one component for $n>1$ or $r>1/2$ modes, though they do
remove two from the $n=1$ and $r=1/2$ modes. Thus if we literally
applied the construction to this case we would arrive at
\bea
\Tr_{\rm Phys} w^R &=&(1-w)^2\prod_{n=1}^\infty (1-w^n)^{-D}
\label{naiveDbraned}\\
\Tr_{\rm Phys} w^R(\pm)^{2R} &=& (1\mp w^{1/2})^2\prod_{r=1/2}^\infty
(1\pm w^r)^{D}\prod_{n=1}^\infty (1-w^n)^{-D}
\label{anivensDbraned}
\eea
instead of (\ref{openpartd}) or (\ref{nsopenpartd}). However, notice
that as far as the $a_1$ and $b_{1/2}$ modes are concerned the
state counting does agree. In particular the massless states still
have only transverse degrees of freedom.

Fortunately there is a deformation of the Virasoro
generators, discovered by David Fairlie 
and me independently in 1971 (see \cite{fairliethorn}), with
an easy extension to the NS super Virasoro generators:
\bea
L^\alpha_n = i\alpha na^{D+1}_n+{\hat L}_n,\qquad G^\alpha_r
=2i\alpha rb^{D+1}_r+{\hat G}_r,
\qquad
L^\alpha_0={\alpha^2\over2}+{\hat L}_0 ,
\eea
where $a^{D+1}_n, b^{D+1}_r$ are the bose and fermi oscillators associated with
the extra dimension, and the hatted generators are
the usual flat space generators in $D+1$ dimensions. These modified
operators satisfy the (super)Virasoro algebra with $c=D+1+12\alpha^2$
for the bosonic open string or
$c=D+1+8\alpha^2$ for the NS open string. 
This deformation can be employed to obtain 
$c=26$ or $c=10$ which then determines $\alpha^2=(25-D)/12$ 
or $\alpha^2=(9-D)/8$. The conformal block built on the
primary state $\ket{0,p}$
will have the degeneracy of open string longitudinal modes,
provided $L_{-1}\ket{0,p}=0$ (or $G_{-1/2}\ket{0,p}=0$ in the NS case).
This determines $\sqrt{2\alpha^\prime}p=i\alpha$.
We must also deform the $J_\pm$ in order that they commute
with the deformed (super)Virasoro operators:
\bea
J_{\pm}&\to&\oint {dz\over2\pi i z}
:e^{i\gamma_\pm y^{D+1}(z)/\sqrt{\alpha^\prime}}:,
\eea
which is well defined on states with 
$p_0\in(\mathbb{Z}/\gamma_\pm-\gamma_\pm)/(2\sqrt{\alpha^\prime}).
$
The condition that $J_\pm$ commute with the Virasoro algebra is that
the vertex operator have conformal weight 1:
\bea
\gamma^2+i\sqrt{2}\alpha\gamma=1,\qquad \gamma_\pm=-i{\alpha\over\sqrt{2}}
\pm\sqrt{1-{\alpha^2\over2}}=-i\sqrt{25-D\over24}\pm\sqrt{D-1\over24}.
\eea
The condition that $J_\pm$ is well defined now reads
\bea
p_0\in -{\gamma_\mp\mathbb{Z}+\gamma_\pm\over2\sqrt{\alpha^\prime}}
\eea
and can be met for both $J_+$ and $J_-$ only for the element $1\in\mathbb{Z}$,
in which case $p_0=i\alpha/\sqrt{2\alpha^\prime}$. It is then very
easy to show that $J_\pm\ket{0,i{\alpha\over\sqrt{2\alpha^\prime}}}=0$.
Then $J_\pm$ kills the whole conformal block built on this ket,
which as noted above is also killed by $L_{-1}$. 
However the algebra of the $J_{\pm}, p_0$ is not the simple SU(2)
we found for the undeformed case.  

In the NS case the deformation that makes $J_\pm$ commute with the
super-Virasoro algebra  reads
\bea
J_{\pm}&\to&\sqrt{2}\oint {dz\over2\pi i z}H(z)
:e^{i\gamma_\pm y^{D+1}(z)/\sqrt{\alpha^\prime}}:,
\eea
which is well defined on states with 
$p_0\in((\mathbb{Z}+1/2)/\gamma_\pm-\gamma_\pm)/(2\sqrt{\alpha^\prime}).
$
The condition that $J_\pm$ commute with the super Virasoro algebra is that
the exponential vertex operator have conformal weight 1/2:
\bea
\gamma^2+i\sqrt{2}\alpha\gamma={1\over2},\qquad 
\gamma_\pm=-i{\alpha\over\sqrt{2}}
\pm\sqrt{{1\over2}-{\alpha^2\over2}}=-i\sqrt{9-D\over16}\pm\sqrt{D-1\over16}.
\eea
The condition that $J_\pm$ is well defined now reads
\bea
p_0\in -{2\gamma_\mp(\mathbb{Z}+1/2)+\gamma_\pm\over2\sqrt{\alpha^\prime}}
\eea
and can be met for both $J_+$ and $J_-$ only for the element $0\in\mathbb{Z}$,
in which case $p_0=i\alpha/\sqrt{2\alpha^\prime}$. It is then very
easy to show that $J_\pm\ket{0,i{\alpha\over\sqrt{2\alpha^\prime}}}=0$.
Then $J_\pm$ kills the whole superconformal 
block built on this ket,
which as noted above is also killed by $G_{-1/2}$. 
But again the algebra of the $J_{\pm}, p_0$ is not the simple O(3)
we found for the undeformed NS case.  

An unappealing feature of the $D+1<$ critical case is the fact that
the values of $p_0$ for which the $J_\pm$ are well-defined are
necessarily complex. This poses conceptual difficulties for
the interpretation of the open string as a unitary worldsheet system
with $p_0^\dagger=p_0$,
because it forces an asymmetry between the kets and bras of
the quantum mechanical interpretation:
\bea
p_0\ket{0,p}=p\ket{0,p}\quad \Longrightarrow \quad \bra{0,p}p_0=p^*\bra{0,p}
\eea
So the dual to $\ket{0,p}$ must be $\bra{0,p^*}$:
\bea
\VEV{0,p^*|0,p}\neq 0, \qquad {\rm but} \qquad \VEV{0,p|0,p}=0.
\eea
On the other hand, this asymmetry fits neatly in a string field 
\cite{thornsft} description as
explained in \cite{preitschopf}. This is because string fields
are naturally associated with the kets, $\ket{A},\ket{A}\otimes\ket{A},\cdots$ 
whereas the bras are used to define terms in the action.

\section{String duals for gauge theory in 4 Dimensions}
The concept of nonabelian D-branes extends the range of
possible string duals for QCD. Since the $\alpha^\prime\to 0$
limit removes all massive string states from the theory, it
is obvious that there are many choices one can make for the
$\alpha^\prime\neq 0$ open string theory. The 4 dimensional NS+ model advocated
in \cite{thornsubqcd} has the simplest mass spectrum. 
For this model, though, we must await better understanding of the
closed string system that couples to this subcritical open
string system.

In the meantime, we can consider open string systems that couple
to critical closed strings. Indeed the original 
${\cal N}=4$/AdS$_5\times$S$^5$ duality is of this type. In this
case the relevant $\alpha^\prime\neq0$ open string system
is just the GSO \cite{gliozziso} projected Neveu-Schwarz-Ramond
\cite{neveuschwarz,neveust,ramond} open string
in 10 space-time dimensions with ends fixed to a stack of
normal (abelian) D3-branes.
We can also consider the NS+ open string model in 10
spacetime dimensions which, together with its
corresponding closed string system, has been called
the type 0 string theory \cite{thornsantafe}.
Adding ordinary D3-branes to this type 0 theory provides a string dual for
Yang-Mills interacting with 6 adjoint massless 
scalars \cite{klebanovtseytlintype0}. However, to reach the gluonic sector
of QCD with this construction, one must find a mechanism to give
large masses to these scalars.

Alternatively, the concept of nonabelian D-branes provides a
way to completely remove these unwanted
massless scalars from the beginning. By imposing
nonabelian Dirichlet conditions, in all six
extra dimensions, on open strings ending on the
D3-branes, one removes (among other things) 
all 6 massless adjoint scalars from the type-0
theory. In this way,
one produces a new candidate for a string dual of large N QCD,
which avoids the conceptual difficulties of the
4 dimensional subcritical NS+ model. The price is
a somewhat richer massive open string spectrum. However, all such
string states become infinitely massive as $\alpha^\prime\to0$, and
so the low energy limit will be pure Yang-Mills gauge theory
with no extra matter fields. For describing QCD, for which
the strong 't Hooft coupling limit is problematic because
of asymptotic freedom, this model 
has the advantage of being cleanly defined as a string theory.

\subsection{Scattering Amplitudes}
The open string tree scattering amplitudes for this critical
NS+ model 
are a subset of the critical NS tree amplitudes: one simply
restricts the external strings to be in  
even G-parity states invariant under
all 6 SU(2)'s associated with the 6 extra dimensions.
Then internal poles of the tree amplitudes factorize only on
this subset of invariant states. In particular, the massless
state (in Picture 2 \cite{neveust}) $b_{-1/2}^I\ket{0,p}$, 
with $I=4,5,\ldots,9$, $p^2=0$, transforms as a
component\footnote{The other
two components of the vector have $p^I\neq0$.} of a vector
 under the $I$th SU(2) and so will decouple. 
Any number of strings in invariant states will never produce a single string
in a noninvariant state.

Of course, NS+ trees with a noninvariant
state on two or more external legs can be nonvanishing. This
means that an amplitude with one or more loops will
differ from the corresponding NS+ amplitude, because every line
participating in a loop must include a projector onto invariant states.
Dealing with these projectors at general loop order is a
challenging unsolved problem. However, at one loop it is not
difficult, especially if the external legs are states, such
as the massless gluon, with no excitations
in the 6 extra dimensions. In that case the modification resides
solely in the partition function factor of the loop integrand. Only one of the
lines in the loop needs to carry the projector, whose effect is 
simply to multiply the usual NS partition function by
the factors $(1\mp w^{1/2})^6$, one factor for each extra dimension
(see Eq~(\ref{nspart})).
In addition, one must remember that, just as with normal
D-branes, the loop momentum integral
is only over the 4 components parallel to the D3-brane.

After the Jacobi transformation to cylinder variables $\ln q
=2\pi^2/\ln w$,
these differences produce the extra factors
\bea
\left[\sqrt{-\pi\over\ln q}(1\mp w^{1/2})\right]^6
=\cases{\left[\int d\mu q^{\mu^2/4}\sin^2{\mu\over2\sqrt{2}}\right]^6\cr
\cr\left[\int d\mu q^{\mu^2/4}\cos^2{\mu\over2\sqrt{2}}\right]^6\cr}
\eea
relative to the usual one loop integrand in $D=10$. 
On the right we have shown the interpretation of the extra factors
as an integral over closed string ``momenta'' $\mu^I$ transverse to
the D3-branes, $I=4,5\ldots,9$. Notice that the normal D-brane would not have
the $(1\mp w^{1/2})^6$ factor, so the $\sin^2$ or $\cos^2$ factor would be
replaced by $1/2$. In these coordinates we see how the
closed string couples to the nonabelian D3-brane. The difference
from the normal D-brane is just the zero mode wave function
factor $\psi_\pm^2(\mu^I)$ 
with $\psi_+(\mu^I)=2^3\prod_{I=4}^9\sin(\mu^I/2\sqrt{2})$
($\psi_-(\mu^I)=2^3\prod_{I=4}^9\cos(\mu^I/2\sqrt{2})$). 
The $+$ and $-$ terms of the loop integrand describe the 
NS-NS and R-R sectors, respectively, of the type-0 closed string system.
The presence of the factor $\psi_\pm$ for a nonabelian D-brane
introduces an asymmetry in the coupling of these two sectors
compared to the normal D-brane. For example at small $\mu$,
corresponding to large distances transverse to the D3-branes,
the NS-NS sector is relatively suppressed.
The closed string system
that propagates in the 10 dimensional bulk is just the
critical type-0 closed string, which is well understood---at least in
perturbation theory. However, as we have just seen, the coupling of these
closed strings to a nonabelian D3-brane is significantly
different from the coupling to the normal (abelian) D3-brane.

\subsection{Lightcone gauge}
We have described a number of open string models
whose $\alpha^\prime\to0$ limit is Yang-Mills in 
weak coupling perturbation theory. To reach $N=\infty$ QCD,
one next has to sum the planar diagrams. Lightcone
quantization of the string \cite{goddardgrt} using
the path integral description \cite{mandelstamlc} provides
one systematic approach to this problem, which promises to be
very useful if ever a numerical assault on the problem
is considered. We close by considering how the nonabelian
D-brane concept fits into the lightcone description.

\begin{figure}[ht]
\begin{center}
\includegraphics[width=1.5in,height=3in,angle=90]{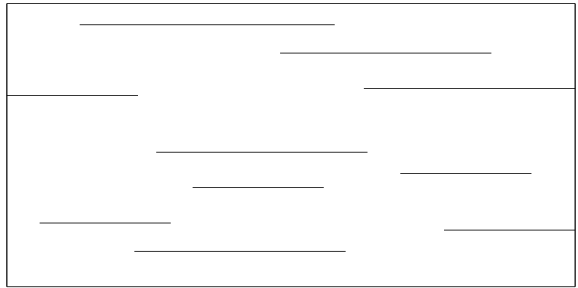}
\caption{A planar lightcone interacting open string diagram with 7 loops.}
\label{lcdiagram}
\end{center}
\end{figure}

A typical multi-loop planar 
lightcone interacting string diagram \cite{mandelstamlc} is shown in
Fig.~\ref{lcdiagram}.
The horizontal dimension of this two dimensional diagram is lightcone
time $x^+=(t+z)/\sqrt2$ and the vertical dimension is $p^+=(p^0+p^z)/\sqrt2$.
The diagram describes the evolution in time of a system of open strings,
breaking and rejoining as shown by the horizontal lines.
For the critical open string, it represents a worldsheet 
path integral whose action is just the
lightcone action for the free open string of interest. To sum
all planar diagrams one simply sums over the
number, length, and location of horizontal lines. 
For each beginning and end of a horizontal line there is a factor
of string coupling $g$, and perhaps a prefactor (as needed for the
NS string). As shown in \cite{bardakcit}
one can mark the presence or absence of a horizontal line at any
point by an Ising spin variable $S(\sigma,\tau)=\pm1$, and execute the
sum as a sum over all spin configurations, with spin dependent
terms in the action enforcing the necessary boundary conditions.

For this interpretation to succeed it is essential that the
path integrand be local. At first glance incorporating 
D-brane constraints seems to spoil locality. However normal D-brane
conditions are not a problem because they are applied
locally on the worldsheet $x_{\rm ends}=0$. But this condition
looks non-local in T-dual variables $\int d\sigma {\dot y}=0$.
Since we only know the nonabelian D-brane conditions in
these T-dual coordinates, we need to consider them a little
more carefully. Each intermediate open string must be projected
onto states invariant under the SU(2)'s for all 
dimensions transverse to the brane. We can express the projector
for the $I$th SU(2) as a group integral
\bea
P_I=\int dR e^{i\theta_a J^a_I},\qquad J_I^a=\int d\sigma {\cal J}_I^a(\sigma)
\eea
For concreteness let's focus on the critical NS+ 
open string ending on nonabelian D3-branes, so the symmetry
is O(3), and $dR$ is the O(3) invariant Haar measure. 
The number of such projector $P=\prod_{I=4}^9 P_I$
insertions changes with the appearance or disappearance of
a horizontal line. Since $P^2=P$ we can put a projector
on each time-slice of each individual string propagator, and introduce
also independent $R$'s for each spacetime point $\sigma, t$:
\bea
P_I=\prod_t P_I=\int \prod_t dR(t)e^{i\theta_a(t)J_I^a}
=\int \prod_t\prod_\sigma dR(\sigma,t)e^{i\int d\sigma\theta_a(\sigma,t){
\cal J}_I^a}
\delta(R^\prime(\sigma,t))\eea
Then by deleting the $\delta(R^\prime(\sigma,t))$ factor whenever
$\sigma$ sits on a horizontal line, we arrive at a worldsheet
local prescription for inserting projectors. In effect we are
gauging each O(3) symmetry on the worldsheet, but replacing
the factor $e^{-\int F^2/4}$ with $\prod_{\sigma,t}\delta(F(\sigma,\tau))$.

One application of this lightcone treatment of the NS+ open string
ending on nonabelian D3-branes is to provide a good regularization
for the lightcone worldsheet construction of the planar diagram
sum in quantum field theory \cite{bardakcit}.
The one loop analysis of that construction for QCD, given
in \cite{chakrabarti1}, showed the need for 
counterterms, whose coefficients could only be determined
by hand. We hope that replacing the field theory planar graphs
with $\alpha^\prime\neq0$ open string graphs as described here
and using the GNS regularization \cite{GNS} 
will systematically determine these counter terms.

\vskip14pt

\noindent\underline{Acknowledgments}: 
I should like to thank Charles Sommerfield for critically
reading the manuscript.
This research was supported in part by the Department
of Energy under Grant No. DE-FG02-97ER-41029.

\end{document}